\documentclass[%
reprint,
showpacs,preprintnumbers,
amsmath,amssymb,
aps,
prb,
]{revtex4-1}

\usepackage{graphicx}% Include figure files
\usepackage{dcolumn}% Align table columns on decimal point
\usepackage{bm}% bold math
\usepackage{colortbl}
\begin{document}
%-------------------------------------------------

%---------------------------------------------------

\title{Kondo correlations formation and the local magnetic moment dynamics in the Anderson model}

\author{N.\,S.\,Maslova$^{1}$}
\altaffiliation{}
\author{P.\,I.\,Arseyev$^{2}$}
\author{V.\,N.\,Mantsevich$^{1}$}
\altaffiliation{} \email{vmantsev@gmail.com}

\affiliation{%
$^{1}$Moscow State University, 119991 Moscow, Russia, $^{2}$ P.N.
Lebedev Physical Institute RAS, 119991 Moscow, Russia
}%

\date{\today }
\begin{abstract}
We investigated the typical time scales of the Kondo correlations
formation for the single-state Anderson model, when coupling to the
reservoir is switched on at the initial time moment. The influence
of the Kondo effect appearance on the system non-stationary
characteristics was analyzed and discussed.
\end{abstract}

\pacs{73.63.Kv, 72.10.Fk, 72.15.Qm} \maketitle

\section{Introduction}

Non-stationary effects now a days attract much attention and are
vital both from fundamental and technological points of view. First
of all, non-stationary characteristics provide more information
about the properties of nanoscale systems comparing to the
stationary ones. Moreover, modern electronic devices design with
particular set of transport parameters requires careful analysis of
non-stationary effects, transient processes and time evolution of
charge and spin states prepared at the initial time moment
\cite{Bar-Joseph},\cite{Gurvitz_1},\cite{Arseyev_1},\cite{Stafford_1},\cite{Hazelzet},\cite{Cota}.

Correct analysis of the non-stationary dynamics of $"$local$"$
magnetic moment and electron occupation numbers of the correlated
Anderson impurity coupled to reservoir requires the investigation of
the Kondo correlations influence on the system time evolution. It is
necessary to clarify the question how the relaxation rates of
$"$local$"$ magnetic moment and charge density change with the
appearance of the Kondo correlations. One can distinguish two main
problems. The first one is widely discussed in the literature and
deals with the Kondo correlations decay (correlations already exist
at the initial time moment) due to the inelastic processes connected
with the many-particle interaction, external field and so on
\cite{Kaminski},
\cite{Rosch_2},\cite{Rosch_1},\cite{Konig},\cite{Nordlander}. In
such situation the typical rate, when the Kondo correlations
disappear is usually connected with the inverse decoherence time
$\tau_{\varphi}^{-1}$. Dephasing rate caused by the inelastic
electron-electron scattering was analyzed in \cite{Kaminski}.
Authors obtained the dependence of spin-flip rate on transferred
energy in two limiting cases: the temperature is higher than the
Kondo temperature and much lower than the Kondo temperature.
Non-equilibrium decoherence rate induced by the voltage driven
current in quantum dot systems was analyzed in \cite{Rosch_2}. The
authors have demonstrated that in the regime of large voltage
(higher than the Kondo temperature) tunneling current prevents the
development of the Kondo correlated singlet state and have found
decoherence rate induced by applied voltage. Later, the dependence
of typical spin-flip rate on the value of external magnetic field in
non-equilibrium case was investigated in \cite{Rosch_1}. The authors
demonstrated that inelastic processes associated with the finite
current through the dot result in the spin-flip effects with typical
rate determined by the renormalized exchange energy. So, tunneling
conductivity and magnetization were found to be universal functions
of $eV/T_{Kondo}$ and $B/T_{Kondo}$, where $eV$ - ia the applied
bias voltage, $B$ - external magnetic field and $T_{Kondo}$ - is the
equilibrium Kondo temperature. The decay rate of the Kondo
correlated state due to photon assisted processes was analyzed in
\cite{Nordlander}, \cite{Kaminski}. It was demonstrated in
\cite{Kaminski} that the dot driven out of equilibrium by an ac
field is also characterized by universal behavior: the dot's
properties depend on the ac field only through the two dimensionless
parameters, which are the frequency and the amplitude of the ac
perturbation, both divided by $T_{Kondo}$.

Another problem which deserves careful analysis deals with the
investigation of the Kondo correlations appearance rate, when
coupling to the reservoir is switched on at the initial time moment
(in such situation the Kondo correlations and any correlations
between the localized and reservoir states are initially absent).
So, the present paper is devoted to the investigation of the typical
time scales responsible for the Kondo correlations formation and the
influence of the Kondo effect on the system non-stationary
characteristics. We show that in the non-stationary case there
exists the only one time scale $T_{Kondo}^{-1}$, responsible for the
formation of the Kondo correlations, which are initially absent.

\section{Theoretical model}

\begin{figure}
\centering
\includegraphics[width=85mm]{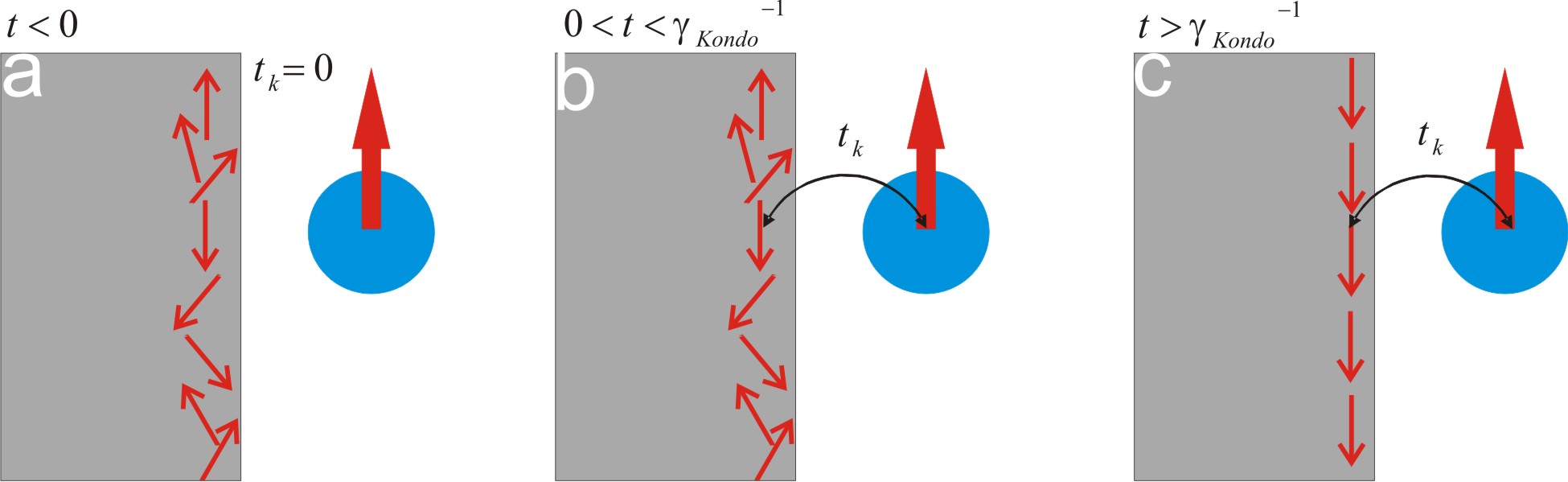}%
\caption{Fig.1 (Color online) Sketch of the Kondo correlations
formation in the proposed system. } \label{figure1}
\end{figure}

We consider non-stationary processes in the system of the
single-level impurity coupled to an electronic reservoir with the
Coulomb interaction of the localized electrons (see
Fig.\ref{figure1}). The model Hamiltonian has the form:

\begin{eqnarray}
\hat{H}&=&\sum_{\sigma}\varepsilon_{1}\hat{n}_{1\sigma}+\sum_{k\sigma}\varepsilon_{k}\hat{c}_{k\sigma}^{+}\hat{c}_{k\sigma}+\nonumber\\
&+&U\hat{n}_{1\sigma}\hat{n}_{1-\sigma}+
\sum_{k\sigma}t_{k}(\hat{c}_{k\sigma}^{+}\hat{c}_{1\sigma}+\hat{c}_{1\sigma}^{+}\hat{c}_{k\sigma}).
\end{eqnarray}

Index $k$ labels continuous spectrum states in the lead, $t_{k}$-
tunneling transfer amplitude between the continuous spectrum states
and localized state with the energy $\varepsilon_1$. $t_{k}$ is
considered to be independent on momentum and spin. Operators
$\hat{c}_{k\sigma}^{+}/\hat{c}_{k\sigma}$ correspond to the
electrons creation/annihilation in the continuous spectrum states
$k\sigma$.
$\hat{n}_{1\sigma(-\sigma)}=\hat{c}_{1\sigma(-\sigma)}^{+}\hat{c}_{1\sigma(-\sigma)}$-localized
state electron occupation numbers, where operator
$\hat{c}_{1\sigma(-\sigma)}$ destroys electron with spin
$\sigma(-\sigma)$ on the energy level $\varepsilon_1$. $U$ is the
on-site Coulomb repulsion for the double occupation of the localized
state. Our investigations deal with the low temperature regime when
Fermi level is well defined and the temperature is much lower than
all the typical energy scales in the system. Consequently the
distribution function of electrons in the leads (band electrons) is
close to the Fermi step.

We are interested in the system dynamics, when coupling to the
reservoir is switched on at the initial time moment. So, any
correlations between localized and reservoir states are not present
initially. Let us consider $\hbar=1$ elsewhere. Kinetic equations
for the electron occupation numbers operators have the form:

\begin{eqnarray}
\frac{\partial \hat{n}_{1\sigma}}{\partial
t}=\sum_{k}t_k(\hat{n}_{1k\sigma}-\hat{n}_{k1\sigma}), \label{0}
\end{eqnarray}

Previously we analyzed long living magnetic moments time evolution
for deep impurities ($\varepsilon_1/\Gamma>>1$) and demonstrated
that for the $"$paramagnetic$"$ initial conditions
($n_{1\sigma}(0)=n_{1-\sigma}(0)$) relaxation rate to the stationary
state is determined by $|\lambda_2|\sim2\Gamma$ and in the case of
the $"$magnetic$"$ initial conditions
($|n_{1\sigma}(0)-n_{1-\sigma}(0)|\sim1$) relaxation rate to the
stationary state is determined by
$|\lambda_1|=\Gamma^{2}/2\varepsilon_1$, where
$\Gamma=\pi\nu_0t_{k}^{2}$ \cite{Mantsevich}. Consequently, the long
living $"$magnetic$"$ moments are present in the system.

\begin{eqnarray}
n_{1\sigma}(t)-n_{1-\sigma}(t)=[n_{1\sigma}(0)-n_{1-\sigma}(0)]\cdot
e^{-\lambda_1t} \label{01}
\end{eqnarray}

We assumed that the Kondo correlations are absent at the initial
time moment are not significant, because they evolve much slower,
than the magnetic moment relaxes. Here arises an important question:
what is the typical rate of the Kondo correlated state formation,
which initially doesn't exist in the system. This rate can be quite
different from the characteristic decay rate of the Kondo
correlation, which were initially present in the system. Another
important problem we are interested in is how the Kondo correlations
appearance influence on the local magnetic moment dynamics. So, we
try to clarify how the Kondo correlations reveal in long living
magnetic moment relaxation and how these correlations evolve with
time.

For simplicity we'll analyze these problems in the case of
infinitely large Coulomb correlations $U\rightarrow\infty$. Then

\begin{eqnarray}
\frac{\partial \langle\hat{n}_{1\sigma}\rangle}{\partial
t}&=&\Gamma\cdot (\int \frac{\langle(1-\hat{n}_{1-\sigma})\hat{n}_{k\sigma}\rangle}{\varepsilon_1-\varepsilon_k+i\Gamma} d\varepsilon_k - h.c.)-2\cdot\Gamma\langle\hat{n}_{1\sigma}\rangle=\nonumber\\
&=&\Gamma\cdot(\int d\varepsilon_k
[\frac{\delta\langle(1-\hat{n}_{1-\sigma})\hat{n}_{k\sigma}\rangle}{\varepsilon_1-\varepsilon_k+i\Gamma}+\nonumber\\
&+&\frac{\langle(1-\hat{n}_{1-\sigma})\rangle\langle\hat{n}_{k\sigma}\rangle}{\varepsilon_1-\varepsilon_k+i\Gamma}]-h.c.)-2\cdot\Gamma\langle\hat{n}_{1\sigma}\rangle
\label{1}
\end{eqnarray}

where
$\delta\langle(1-\hat{n}_{1-\sigma})\hat{n}_{k\sigma}\rangle=\langle(1-\hat{n}_{1-\sigma})\hat{n}_{k\sigma}\rangle-\langle(1-\hat{n}_{1-\sigma})\rangle\langle\hat{n}_{k\sigma}\rangle$.
The first term describes the corrections to the system dynamics
beyond slowly varying amplitudes approximation. Operator equations
of motion have the following form:

\begin{eqnarray}
\frac{\partial[(1-\hat{n}_{1-\sigma})\hat{n}_{k\sigma}]}{\partial
t}&=&-t_{k}\sum_{k^{'}}(\hat{n}_{1k^{'}-\sigma}-\hat{n}_{k^{'}1-\sigma})\hat{n}_{k\sigma}-\nonumber\\
&-&t_{k}(1-\hat{n}_{1-\sigma})(\hat{n}_{1k\sigma}-\hat{n}_{k1\sigma}),\nonumber\\
\langle\hat{n}_{k\sigma}\rangle\frac{\partial
\langle(1-\hat{n}_{1-\sigma})\rangle}{\partial
t}&=&-t_{k}\sum_{k^{'}}\langle\hat{n}_{1k^{'}-\sigma}-\hat{n}_{k^{'}1-\sigma}\rangle\langle\hat{n}_{k\sigma}\rangle\nonumber\\
\label{2}
\end{eqnarray}

Let's consider equation for $\hat{n}_{1k\sigma}$. After averaging
over fast oscillations it can be expressed as:

\begin{eqnarray}
\hat{n}_{1k\sigma}&=&\frac{t_{k}(1-\hat{n}_{1-\sigma})(\hat{n}_{1\sigma}-\hat{n}_{k\sigma})}{\varepsilon_1-\varepsilon_k+i\Gamma}+\nonumber\\
&+&t_{k}\sum_{k^{'}}\frac{\hat{n}_{1k^{'}-\sigma}\hat{n}_{1k\sigma}}{\varepsilon_1-\varepsilon_k+i\Gamma}-(k^{'}\leftrightarrow1)\label{3}
\end{eqnarray}

Applying sequential iteration procedure one can obtain:

\begin{eqnarray}
\hat{n}_{1k\sigma}&=&\frac{t_{k}(1-\hat{n}_{1-\sigma})(\hat{n}_{1\sigma}-\hat{n}_{k\sigma})}{\varepsilon_1-\varepsilon_k+i\Gamma}+\nonumber\\
&+&t_{k}^{2}\sum_{k^{'}}\hat{n}_{1k^{'}-\sigma}\frac{(1-\hat{n}_{1-\sigma})(\hat{n}_{1\sigma}-\hat{n}_{k\sigma})}{(\varepsilon_1-\varepsilon_k+i\Gamma)^{2}}+\nonumber\\
&-&t_{k}^{4}\sum_{k^{'}k^{''}}\hat{n}_{k^{'}-\sigma}\hat{n}_{1k^{''}-\sigma}\frac{(1-\hat{n}_{1-\sigma})(\hat{n}_{1\sigma}-\hat{n}_{k\sigma})}{(\varepsilon_1-\varepsilon_k^{'}+i\Gamma)(\varepsilon_1-\varepsilon_k+i\Gamma)^{3}}+...\nonumber\\
\label{3}
\end{eqnarray}

Let us now analyze the nontrivial logarithmic divergent terms, which
appear beyond the decoupling approximation.

\begin{eqnarray}
t_{k}^{2}\sum_{k^{'}}\hat{n}_{1k^{'}-\sigma}\frac{(1-\hat{n}_{1-\sigma})(\hat{n}_{1\sigma}-\hat{n}_{k\sigma})}{(\varepsilon_1-\varepsilon_k+i\Gamma)^{2}}\sim\nonumber\\
\sim t_{k}\frac{(1-\hat{n}_{1-\sigma})(\hat{n}_{k\sigma})}{\varepsilon_1-\varepsilon_k+i\Gamma}\cdot \sum_{k^{'}}\frac{(1-\hat{n}_{1\sigma})\langle\hat{n}_{k^{'}-\sigma}\rangle\cdot t_{k}^{2}}{(\varepsilon_1-\varepsilon_k^{'}+i\Gamma)(\varepsilon_1-\varepsilon_k+i\Gamma)}\sim\nonumber\\
\sim\frac{\Gamma\cdot
ln(\varepsilon_k/W)}{\varepsilon_1-\varepsilon_k+i\Gamma}\cdot
t_{k}\frac{(1-\hat{n}_{1-\sigma})(\hat{n}_{k\sigma})}{\varepsilon_1-\varepsilon_k+i\Gamma}+...\nonumber\\
\label{5}
\end{eqnarray}

Here we used the following relation
$\hat{n}_{1\pm\sigma}(1-\hat{n}_{1\pm\sigma})=0$. Retaining the most
divergent logarithmic terms in the higher order iterations, we
obtain:

\begin{eqnarray}
\hat{n}_{1k\sigma}\sim\frac{t_{k}(1-\hat{n}_{1-\sigma})(-\hat{n}_{k\sigma})}{\varepsilon_1-\varepsilon_k+i\Gamma}\cdot\nonumber\\
(1+\frac{\Gamma\cdot
ln(\varepsilon_k/W)}{\varepsilon_1-\varepsilon_k+i\Gamma}+[\frac{\Gamma\cdot
ln(\varepsilon_k/W)}{\varepsilon_1-\varepsilon_k+i\Gamma}]^{2}+.....)\sim\nonumber\\
\sim(\varepsilon_k<<\Gamma)\sim-\frac{t_{k}(1-\hat{n}_{1-\sigma})(\hat{n}_{k\sigma})}{\varepsilon_1(1-\frac{\Gamma}{\varepsilon_1}\cdot
ln(\varepsilon_k/W)+i\frac{\Gamma}{\varepsilon_1})}+... \label{6}
\end{eqnarray}

Introducing $\gamma_{Kondo}=We^{\frac{-|\varepsilon_1|}{\Gamma}}$ in
the usual way, one can easily get:

\begin{eqnarray}
t_{k}\hat{n}_{1k\sigma}=-\frac{\Gamma}{\nu_0\varepsilon_1}\cdot\frac{(1-\hat{n}_{1-\sigma})(\hat{n}_{k\sigma})}{\frac{\Gamma}{\varepsilon_1}\cdot
ln(\varepsilon_k/\gamma_{Kondo})+i\frac{\Gamma}{\varepsilon_1})}+\nonumber\\+\frac{\Gamma}{\nu_0}\frac{(1-\hat{n}_{1-\sigma})(\hat{n}_{1\sigma}-\hat{n}_{k\sigma})}{(\varepsilon_1-\varepsilon_k+i\Gamma)}
\label{7}\end{eqnarray}

Finally, substitution of expression (\ref{7}) to Eq.(\ref{2}) yields

\begin{eqnarray}
\frac{\partial}{\partial
t}\delta\langle(1-\hat{n}_{1-\sigma})\hat{n}_{k\sigma}\rangle=\nonumber\\=-\gamma_{Kondo}\langle\delta(1-\hat{n}_{1-\sigma})\hat{n}_{k\sigma}\rangle+\frac{1}{\nu_0}\frac{\langle(1-\hat{n}_{1-\sigma})\rangle\langle\hat{n}_{k\sigma}\rangle}{ln^{2}(\varepsilon_k/\gamma_{Kondo})+1}\nonumber\\\label{8}
\end{eqnarray}

Let's define
$\sum_{k}\delta\langle(1-\hat{n}_{1-\sigma})\hat{n}_{k\sigma}\rangle=K^{-\sigma\sigma}$.
The main contribution to $K^{-\sigma\sigma}$ comes from
$\varepsilon_k\sim\varepsilon_F$. So, one can obtain:

\begin{eqnarray}
\frac{\partial <n_{1-\sigma}>}{\partial
t}&=&-2\Gamma[<n_{1-\sigma}>-(1-<n_{1\sigma}>)
N_{k\varepsilon}^{-\sigma}-\nonumber\\&-&\frac{\Gamma}{2\varepsilon}\cdot K^{\sigma-\sigma}(t)],\nonumber\\
\frac{\partial <n_{1\sigma}>}{\partial
t}&=&-2\Gamma[<n_{1\sigma}>-(1-<n_{1-\sigma}>)
N_{k\varepsilon}^{\sigma}-\nonumber\\&-&\frac{\Gamma}{2\varepsilon}\cdot K^{-\sigma\sigma}(t)],\nonumber\\
\frac{\partial K^{\sigma-\sigma}(t)}{\partial
t}&=&-\gamma_{Kondo}\cdot[K^{\sigma-\sigma}(t)-K_{0}^{\sigma-\sigma}(t)],\nonumber\\
\frac{\partial K^{-\sigma\sigma}(t)}{\partial
t}&=&-\gamma_{Kondo}\cdot[K^{-\sigma\sigma}(t)-K_{0}^{-\sigma\sigma}(t)]\label{9}\nonumber\\
\end{eqnarray}

where

\begin{eqnarray}
K_{0}^{-\sigma\sigma(\sigma-\sigma)}(t))=N_{Kondo}\cdot\langle(1-n_{1\mp\sigma})\rangle,\nonumber\\
N_{Kondo}=\frac{1}{\pi}\int\frac{\gamma_{Kondo}}{\gamma_{Kondo}^{2}+\varepsilon_{k}^{2}}\cdot
n_{k}(\varepsilon_{k})d\varepsilon_{k}.\label{10}
\end{eqnarray}

and

\begin{eqnarray}
N_{k\varepsilon}^{\sigma}=N_{k\varepsilon}^{-\sigma}=\frac{1}{2\pi}i
\int
d\varepsilon_{k}n_{k\sigma}(\varepsilon_{k})\times\nonumber\\ \times[\frac{1}{\varepsilon_1+i\Gamma-\varepsilon_{k}}-\frac{1}{\varepsilon_1-i\Gamma-\varepsilon_{k}}]\label{11}\nonumber\\
\end{eqnarray}

Initial conditions are $n_{1\pm\sigma}(0)=n_{0\pm\sigma}$ and
$K^{-\sigma\sigma(\sigma-\sigma)}(0))=0$. At $t=0$ the Kondo
correlations and any correlations between the impurity and reservoir
states are absent. So, the characteristic time scale of the Kondo
correlations formation can be defined as $\gamma_{Kondo}^{-1}$. This
time scale of the Kondo correlations formation differs from the
typical time of their decay, caused by the inelastic interaction
(voltage driven current, electron-phonon interaction etc.). The
appearance of the Kondo correlated state is governed by the exchange
interaction between localized and conduction electron in reservoir,
while the decay of the Kondo correlations is determined by inelastic
spin-flip processes with the characteristic rate $\tau_{sf}^{-1}$.
At low temperature $T<\gamma_{Kondo}$ this rate can be determined as
$\frac{1}{\tau_{\phi}}\sim\frac{\omega}{ln^{2}(\frac{\omega}{\gamma_{Kondo}})}$,
where $\omega$ is the typical transferred energy value due to
inelastic interaction. In the case of voltage driven current
$\omega$ has to be replaced by $eV$.

\begin{figure*}
\centering
\includegraphics[width=180mm]{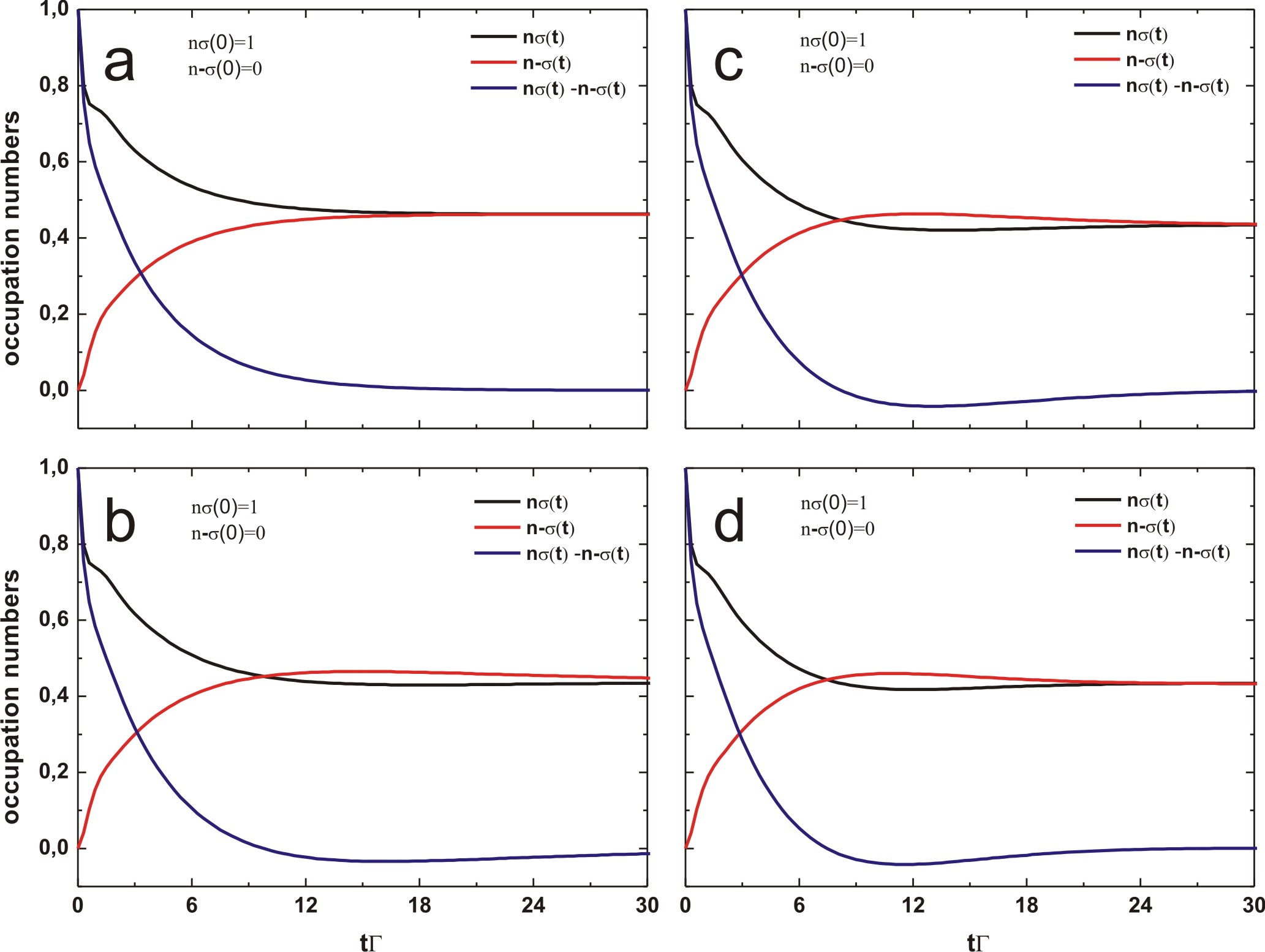}%
\caption{Fig.2 (Color online) Electron occupation numbers
$n_{1\pm\sigma}(t)$ and local magnetic moment
$n_{1\sigma}(t)-n_{1-\sigma}(t)$ time evolution. a).
$\gamma/\Gamma=0.00$; b). $\gamma/\Gamma=0.05$; c).
$\gamma/\Gamma=0.10$; d). $\gamma/\Gamma=0.15$. Parameters
$\varepsilon/\Gamma=-2.5$ and $\Gamma=1$ are the same for all the
figures.} \label{figure2}
\end{figure*}

Obtained results (see Fig. \ref{figure2}) also demonstrate long
living magnetic moment relaxation with typical rate $\lambda_1$ up
to the time $t\sim\gamma_{Kondo}^{-1}$, which is the time of the
Kondo correlations (absent at initial time moment) appearance. For
the times $t>\gamma_{Kondo}^{-1}$ electron occupation numbers time
evolution $n_{1\pm\sigma}(t)$ demonstrate spin-flip effects, caused
by the presence of the Kondo correlations. Spin-flip effects lead to
the non-monotonic behavior of electron occupation numbers and to the
changing of local magnetic moment sign (see Fig. \ref{figure2}). But
this effect is weak, because $\lambda_1$ strongly exceeds
$\gamma_{Kondo}$ and local magnetic moment nearly approaches to it's
stationary value, when the Kondo correlations appear. Slow changing
of local magnetic moment sign near it's stationary value (which is
equal to zero) is clearly seen from Fig. \ref{figure2}. With further
time increasing local magnetic moment reaches it's stationary zero
value for non-magnetic reservoir.

At the large time scales $t>\gamma_{Kondo}^{-1}$ Kondo correlations
lead to slight decreasing of local magnetic moment relaxation rate.
At such time scales the correlation function $K^{-\sigma\sigma}$ is
close to it's stationary value. So, the magnetic moment relaxation
rate $\lambda_1=2\Gamma(1-N_{k\varepsilon}^{\sigma})$ is replaced by

$2\Gamma(1-N_{k\varepsilon}^{\sigma}-\frac{\Gamma}{2\varepsilon}\cdot
N_{Kondo})$.

\section{Conclusion}

We analyzed non-stationary processes of the Kondo correlations
formation, when coupling to reservoir is switched on at the initial
time moment. It was found out that the typical time scale of the
Kondo correlations appearance is $\gamma_{Kondo}^{-1}$, which is
quite different from decoherence time associated with the inelastic
spin-flip processes. The influence of the Kondo effect on the
non-stationary dynamics of local magnetic moment and electron
occupation numbers of the correlated Anderson impurity coupled to
reservoir was investigated.

It was demonstrated that for the times $t>\gamma_{Kondo}^{-1}$
electron occupation numbers time evolution is weakly influenced by
the spin-flip effects, caused by the appearance of the Kondo
correlations, because the relaxation rate of local magnetic moment
strongly exceeds $\gamma_{Kondo}$. It was also revealed that for the
large time scales $t>\gamma_{Kondo}^{-1}$ the Kondo correlations
lead to slight decreasing of local magnetic moment relaxation rate.

This work was supported by RFBR grant $16-32-60024$ $mol-a-dk$.

 \pagebreak

\end{document}